\documentclass{article}
\usepackage{graphicx}

\begin{document}
\rm
\let\sc=\sf

\LARGE
\begin{center}
{\bf The Disk Wind in the Young Binaries and the Origin
of the Cyclic Activity of Young Stars}

\vspace{1cm}

\Large
{\bf V.P.\,Grinin$^{1,2,3}$, L.V.\,Tambovtseva$^1$, N.Ya.\,Sotnikova$^2$}
\end{center}

\vspace{5mm}
\large

1 - Main Astronomical Observatory Pulkovo, 196140, St. Petersburg, Russia

2 - The Sobolev Astronomical Institute, St. Petersburg University, Russia

3 - Crimean Astrophysical Observatory, Crimea, Nauchny, Ukraine\\
\\
\begin{abstract}
\large
\hspace{-5mm} We present results of numerical modeling of
the cyclic brightness modulation in the young binary systems with
the eccentric orbits and low-mass secondary components. It is
suggested that the system components accrete the matter from the
remnant of the protostellar cloud and, according to the current
models, the low-mass companion is the main accretor. Brightness
variations of the primary is due to the periodical extinction
variations on the line-of-sight caused by the disk wind of the
secondary and a common envelope it produces. A matter distribution
in the envelope has been calculated in the ballistic approach.

When calculating the optical effects due to the dust component of
the disk wind, we adopt the dust to gas mass ratio 1:100 as in the
interstellar medium and the optical parameters of the
circumstellar dust typical for the young stars. Calculations
showed that in the young binaries with the elliptic orbits
theoretical light curves demonstrated the more variety of shapes
comparing with the case of the circular orbits. In this case
parameters of the photometric minima (their depth, duration and
the shape of light curves) depend not only on the disk wind
parameters and an inclination of the binary orbit to the
line-of-sight but also on the longitude of the periastron. A
modulation of the scattered radiation of the common envelope with
a phase of the orbital period has been investigated in the single
scattering approach. It is shown that an amplitude of the
modulation is maximal when the system is seen edge-on and has also
a non-zero value in the binaries observed pole-on. Possible
applications of the theory to the young stellar objects are
discussed. In particular, an attention is payed to a resemblance
of the light curves in some models with light curves of the
objects suspected as candidates to FUORs.

\end{abstract}

\clearpage
\section{Introduction}
Photometric observations of the last years show that among young
stars one can find the eclipsing systems with a rather long
lasting eclipses. For example, in the binary system GW Ori (P =
242$^d$, Mathieu et al. 1991) where the primary is a T Tauri star
(TTS), a duration of the eclipses is about 1/10 of the period
(Shevchenko et al. 1998). The duration of the eclipses of the weak
T Tauri star (WTTS) KH 15D (P = 48.36$^d$, Kearns and Herbst 1998;
Hamilton et al. 2001; Herbst et al. 2002) is even more: one third
of the period. In the binary HD 200775, whose period is about of 3
years (Pogodin et al. 2004) and where the main component is the
Herbig Ae/Be star, the duration of the eclipses is comparable with
the orbital period (Ismailov 2003). Not long ago one more WTTS H
187 with the eclipse of 3.6 years has been discovered in the young
cluster IC 348 (Cohen et al. 2003). To date only one eclipse has
been observed completely. Therefore, the period of this system is
estimated rather approximately and according to Cohen et al. is
about of 4 years. An interpretation of such eclipses by classical
models developed for the eclipsing binaries, where either the
secondary itself or the gas and dust envelope filling in its Roshe
lobe is "an obscuring body", leads to the serious contradictions
with the physics of the young stars; in some cases such an
interpretation is impossible in principle as in the case of KH 15D
and H 187, since during such long lasting eclipses the gas and
dust envelope around the secondary must have sizes comparable with
the radius of the orbit. Such envelopes are unstable and are
quickly to destroy due to the tidal perturbations.

Recently it was shown (Grinin and Tambovtseva 2002 (Paper I)) that
in the young binaries which still continue to accrete the matter
from the remnants of the protostellar cloud, a rather unusual
mechanism of eclipses can occur where the role of "the obscuring
body" belongs to the disk wind of the secondary component, namely,
to its dusty component. Because of an extended structure of the
disk wind, the eclipses evoked by it can be very prolonged. Unlike
the classical models of the eclipsing binaries, where the eclipses
occur only when the line-of-sight lies in the orbit plane or
deviates from the latter at a small angle, the eclipses by the
disk wind are possible even at the large inclination angles. The
more the inclination angle the longer can be the eclipses. This
allows us to suppose that a number of the eclipsing binaries among
the young pairs has to be i) larger in comparison with the
binaries of the Main Sequence, and ii) the number of the eclipsing
systems with the long eclipses has to be larger among the young
pairs.

In connection with this, it is interesting to continue a study of
the optical effects caused by the disk wind in the young binaries.
In the present paper which is continuation of the Paper I we give
the results of the numerical simulation of the cyclic phenomena
due to the disk wind of the low mass secondary during its motion
on the orbit. We investigate a dependence of the theoretical light
curves on the disk wind parameters as well as parameters of the
binary including its orientation in the space. Unlike Paper I
where a solution of such a problem has been obtained for the
systems with the circular orbits, here we consider a common case
of the binaries with the elliptic orbits.

\clearpage
\section{The problem}

As in Paper I, we assume the model of the young binary proposed by
Artymowizc and Lubow (1996) (AL96) as the basic one, and restrict
ourselves to the case when the mass of the secondary companion is
much less than that of the primary\footnote{According to the
statistics of the Main-Sequence binaries (Duquennoy and Mayor
1991; Mazeh et al. 2000) such mass component ratios are typical
for the systems with periods $P \ge 3$ yrs but they can be also
found in the systems with shorter periods. In particular, it could
be systems with the substellar companions whose number rapidly
grows owing to the continuing research programmes (Mayor and Urdy
2000).}. In the center of such a system a cavity almost free of
the matter is formed under the effect of the periodic
gravitational perturbation (Fig. 1). Its typical size depends on
the eccentricity $e$ and the mass ratio of the components $q =
m_2/m_1$ and is approximately equal to $(2-3)\cdot a$ where $a$ is
a large semi-axis of the orbit of the secondary (Artymowizc and
Lubow 1994).

The components of the binary accrete the matter from the remnants
of the protostellar cloud which forms so-called circumbinary (CB)
disk. In this case , the accretion rate onto the low mass
component can substantially exceed the accretion rate onto the
primary(Artymowizc and Lubow 1994, Bate and Bonnel 1997; Rozyczka
and Laughlin 1997). Since the process of the disk accretion is
accompanied with the matter outflow from the accretion disk (the
disk wind) then at such conditions, the low mass companion becomes
a powerful source of the matter which it ejects up (and down) the
CB disk plane during its motion along the orbit. Calculations
showed that the dust presented in the wind\footnote {As Safier
(1993) showed, due to collisions of the dust particles with gas
atoms in the accretion disk, the former are carried away by the
latter and are present approximately in the same proportion in the
disk wind.} can originate the long lasting eclipses of the primary
(Paper I) and could be (at the certain conditions) the source of
the thermal radiation whose luminosity in the near infrared region
of the spectrum can be comparable with that of the CB disk itself
(Grinin 2002).

\subsection{A structure of the CB-disk}
As an example, a structure of the CB-disk of the young binary with
the low mass component is shown on Fig. 1. We calculated the
matter distribution in the disk in the hydrodynamical approach by
SPH (smoothed particle hydrodynamics) method using a scheme close
to that described by Hernquist and Katz (1989) but with a constant
smoothing length of the hydrodynamical parameters.
\begin{figure*}
\vspace{-1.cm}
\hspace{-1cm}\includegraphics[width=20cm]{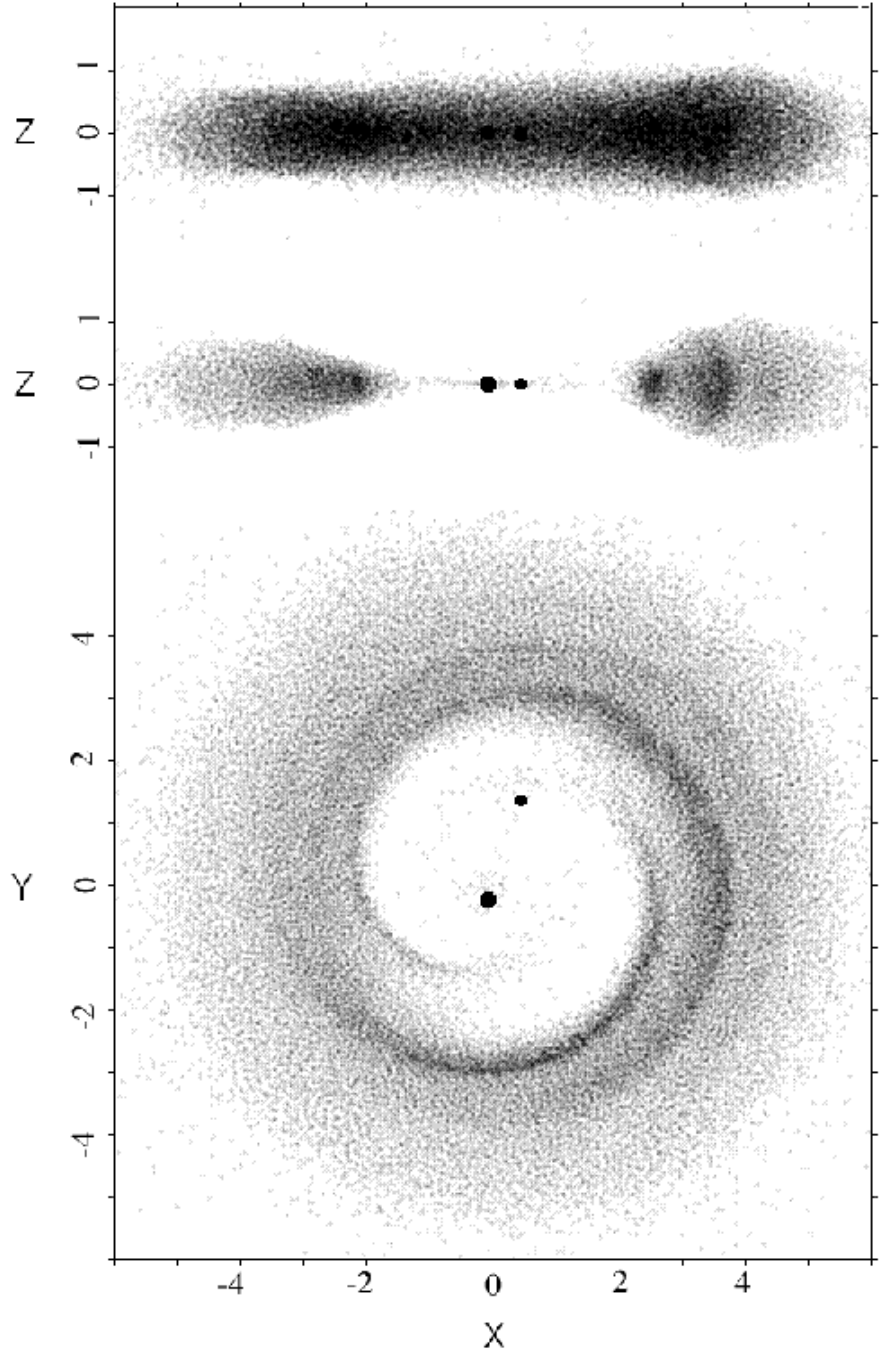}
\vspace{-9cm} \caption{A matter distribution in the CB-disk of the
young binary obtained with SPH-method; the view "edge-on" in the
projection on $XZ$ plane (the upper panel), the cross-section of
the CB-disk in the same plane (the middle panel) and the view
"pole-on" (the bottom). The co-ordinates $X, Y, Z$ are expressed
in the units of the large semi-axis of the secondary's orbit. The
model parameters are $e=0.5$, mass ratio $q = 0.22$}
\end{figure*}
In the projection on the sky plane one can see two streams of the
matter from the CB disk feeding the accretion disks around the
components of the system. In the CB disk itself the wave densities
caused by the periodical gravitational perturbations are seen.
Such a matter distribution in the CB disk agrees well with results
obtained by AL96.

A special feature of the binaries with the elliptic orbits is a
global asymmetry in the azimuthal distribution of the CB disk
matter that is clearly seen both pole-on and edge-on as well as in
the cross-section (Fig. 1). As it was shown by Artymowizc and
Lubow (2000), the asymmetric CB disk precesses slowly with a
precession period significantly exceeding the orbital one. In the
cross-section (Fig. 1, middle panel) CB disk resembles a classical
accretion disk around a single young star in which a main part of
the matter is concentrated in the geometrically thin equatorial
layer with a thickness $H \ll r$.

One of the manifestation of the CB disk global asymmetry mentioned
above is a dependence of its geometrical thickness $H$ not only on
the distance from the center $r$ but also on the azimuth. For this
reason, the precession of the CB disk has to originate a long
lasting periodical variations of the extinction in the young
binaries observed nearly edge-on when the line-of-sight is tangent
to "the surface" of the disk. It is also obviously, that the
global asymmetry of the CB disk is one of the sources of the
intrinsic polarization of the young binaries.

\section{Formation of the common envelope}
The disk wind of the secondary essentially modifies the model of
the young binary: during the orbital motion of the secondary it
creates a rather complex in its structure asymmetric common
envelope (Fig. 2) which partially dissipates in the surrounding
space but partially can be captured by the main component. As a
result, a dust appears above (and under) the binary system's plane
giving an additional extinction on the line-of-sight.
\begin{figure*}
\vspace{-0.5cm}
\hspace{1cm} \includegraphics[width=20cm]{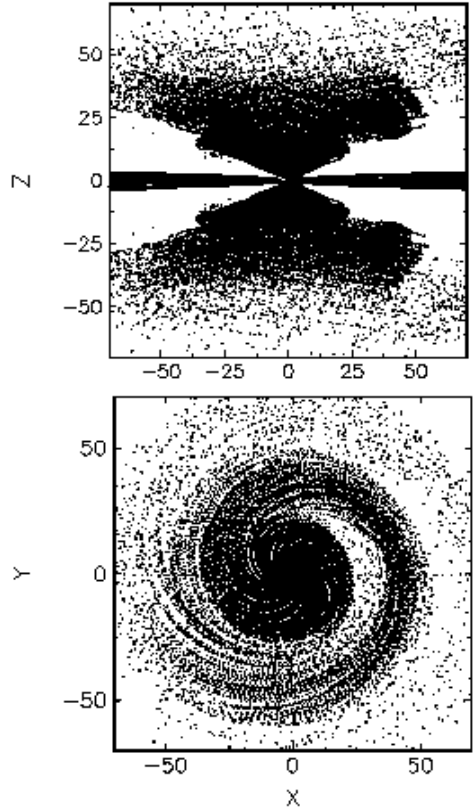}
\vspace{-15cm} \caption{The matter distribution in the common
envelope of the young binary created by the disk wind of the
secondary in the projection on the $XZ$-plane (top) and on the
equatorial $(XY)$ plane (bottom). The secondary is at the point
with the coordinates $X = 1, Y = 0$. The coordinates of the
primary in the $XY$-plane are equal to (0,0). All coordinates are
expresses in the units of the large semi-axis of the secondary's
orbit. The model parameters are $V_w = 3, U_w = 0.5, e = 0.5$ (see
Section 3.2)}.
\end{figure*}

\subsection{Inertial properties of the accretion disks}
A special feature of the binaries with the elliptic orbits is that
the accretion rate onto the system components undergone a strong
modulation with a period equaled to the orbital one (AL96).
Moreover, in the binaries with elongated orbits ($e \ge 0.3$) a
modulation amplitude can be so large that the accretion rate in
the apastron and in the periastron can differ by ten and more
times (AL96; Rozyczka and Laughflin 1997). The dependence of the
accretion rate $\dot M_a$ on the orbital period phase can lead to
that the mass loss rate in the disk wind $\dot M_w$ can be also a
periodic function of time. Nevertheless, due to an inertia of the
accretion disk, the modulation amplitude of $\dot M_w$ depends on
the ratio between the orbital period $P$ and the hydrodynamical
time $t_g$. The latter is a characteristic time of the accretion
disk dissipation without any feeding. If it is less compared to
the orbital period $P$ then the accretion disk has a time to react
to the changes on its outer boundary caused by the changes in the
accretion rate. In this case $\dot M_w \propto \dot M_a$. In the
opposite case, a reaction of the disk to the changes of $\dot M_a$
can be strongly smoothed.

According to Shakura and Sunyaev (1973)
\begin{equation}
t_g = R^2/\nu
\end{equation}

Here $R$ is the outer radius of the accretion disk, $\nu$ is a
coefficient of the turbulent viscosity: $\nu = \alpha \,v_s\,H$,
where $H$ is a geometrical thickness of the disk, $v_s$ is a sound
speed and $\alpha$ is a dimensionless parameter depending on the
mechanism of the generation of turbulence in the accretion disk.
Taking into account that $v_s/v_\phi \approx H/R$ where $v_\phi$
is a Keplerian velocity on the outer boundary of the disk and
expressing $t_g$ in the units of the orbital period we obtain
\begin{equation}
t_g = (2\pi \alpha)^{-1} (R/H)^2 (t_{rot}/P)\,,
\end{equation}
where $t_{rot}$ is a Keplerian period at the outer boundary of the
disk of the low mass companion.

The expression (1) one can rewrite in the form
\begin{equation}
t_g = (2\pi \alpha)^{-1} (R/H)^2 q^{-1/2} (R/a)^{3/2}\,
\end{equation}
where $a$ is the large semi-axis of the secondary, $q =
m_2/m_1$ is the mass ratio of the components.

According to Artymowizc and Lubow (1994), if the mass ratio $q
\approx 0.1$ the outer radius of the secondary accretion disk
$R \approx 0.1\,a$. Assuming a "standard" relation
for the accretion disks $H\,/R$ = 0.05 and $R$ = 0.1\,$a$, we
obtain from Eq. (3) $t_g \approx 6.5\,\alpha^{-1}$.

In the modern models (see the review by Stone et al. 2000 and the
references therein) the values of $\alpha$ are ranging from 0.005
to 0.5. Assuming the value $\alpha = 0.5$ to estimate the lower
limit of $t_g$, we obtain $t_g \ge 13$.

Hence it follows, that even in the models with the high efficiency
of the angular momentum transfer a typical hydrodynamical time of
the accretion disk substantially exceeds the orbital period. From
this one can conclude that the disk wind forming mainly in the
central part of the disk is in no time "to feel" the changes of
the accretion rate at its outer boundary. However, observations of
the young binaries with the eccentric orbits testifies (see e.g.
Pogodin et al. 2004) that there is a modulation of some parameters
of the emission line $H_\alpha$ with the phase of the orbital
period such as an equivalent width of the line and parameters of
its profile. This means, that physical conditions in the central
part of the accretion disk where this line is formed undergone the
periodical variations connected with the orbital motion of the
system components. This variations can be caused by the tidal
perturbations and spiral shock waves forming in the accretion
disks of the binaries (Sawada et al. 1986a,b; Bisikalo et al.
1995; Makita et al. 2000). As a result, the disk wind can be
strengthened near the orbit periastron. Therefore, further
together with the conservative model where $\dot M_w = const$, we
also consider the model of the binary where the disk wind is
strengthening when the system components are approaching each
other. For simplicity we investigated the case when the mass loss
rate is proportional to the accretion rate.

\subsection{Parameters of the disk wind}
We remind briefly the main characteristics of the disk winds in
the young stars. The numerical modeling by Goodson et al. (1999)
shows that the bulk of the matter (up to 80\%) is concentrated in
the low velocity component of the wind and ejected from the
accretion disk in the angle range $\omega \approx 40 ^\circ -
60^\circ$ where $\omega$ is the angle between the vector of the
radial velocity of the wind and the symmetry axis of the disk.
Near the accretion disk, the disk matter has not only the radial
velocity component $V_w$ but also the azimuthal one $U_w$. At the
large distances from the accretion disk the latter decreases due
to the angular momentum conservation law and the radial velocity
becomes the dominant component. Further, putting the kinematic
wind parameters, we shall mean the velocity components $V_w$ and
$U_w$ as those which the wind fragments have after completing the
acceleration phase when the wind motion occurs inertly. Hartigan
et al. (1995) and Hirth et al. (1997) estimated a typical low
velocity component of the wind from the observations of the
forbidden lines in the spectra of TTSs: few tens kilometers per
second at the distances $\ge$ 1 AU.

Taking into account all mentioned above, we assume that all wind
particles are ejected with the same radial velocities $V_w$
isotropically in the angle range $40^\circ  \le \omega \le
60^\circ$. It is supposed that the disk wind possesses a mirror
symmetry relatively to the orbit plane. Since the radius of the
accretion disk of the secondary is less compared to the large
semi-axis then, for simplification of calculations we adopt that
the matter outflows from the point source whose coordinates
coincide with those of the secondary, and the mass center of the
system coincides with the primary location. The disk wind of the
primary was not taking into consideration when modeling because
the accretion rate onto this component is much less than that onto
the secondary according to the problem condition.

It is also suggested that in the coordinate system of the
secondary the disk wind has an axial symmetry. This condition is
fulfilled in those cases when one can neglect the tidal
perturbations due to the main companion. Since the radius of the
accretion disk of the secondary is less compared to the radius of
the tidal interaction and the main contribution to the disk wind
is given by the internal layers of the accretion disk, such a
suggestion seems quite reasonable. An exclusion are binaries with
a very large eccentricity. In such systems the outer radius of the
secondary accretion disk at the orbit periastron can turn out less
than the radius of the tidal interaction; as a result, the part of
the matter from this disk can be captured by the primary at the
moment of the maximal approach. At these conditions one can expect
strong deviations from the azimuthal symmetry in the disk wind.

\section{Method of calculation}
As in Paper I we suppose that the disk wind consists of the weakly
interacting fragments which are treated as independent (probe)
particles. When going to the coordinate system of the primary the
velocity vector of the particle {\bf V}$_w$ + {\bf U}$_w$ is
summed with the velocity vector of the secondary orbital motion
{\bf V}$_k$. As a result the disk wind becomes anisotropic:
\begin{equation}
{\bf V}_0 = {\bf V}_w  + {\bf U}_w + {\bf V}_k.\,
\end{equation}
In this case the part of the matter (ejected in the directions
opposite to that of the orbital motion of the secondary)
can get the velocity less than the escape one and can be captured
by the main component.

The calculation of the trajectories of the particles motion in the
gravitational field of the primary\footnote{Masses of the
circumstellar (CS) disks of the young stars usually do not exceed
0.1\, $M_{\odot}$ (see, e.g. Natta et al. 2000) and their
self-gravity can be neglected.} was carried out in the ballistic
approach. The particle velocity $\bf V_0$ determined by Eq. (4)
and its coordinates at the moment of ejection are put as initial
conditions in calculation of its trajectory. In the computing
process an orbit of the secondary was divided by $n$ fragments in
such a way that its motion along each part of the orbit took the
same time $\Delta t = P/n$. We put $n$ = 72 or 180 (that in the
case of the circular orbit corresponded to the orbit step equal to
5$^\circ$ or 2$^\circ$ correspondingly). The same algorithm of the
common envelope modeling as in Paper I has been used. Model
simulations of the disk wind was carried out via ejection of the
probe particles on each step isotropically within the range of
solid angles mentioned above into the upper and lower semi-space.

Thus, the model parameters of the problem are:
\begin{itemize}
\item an eccentricity of the orbit $e$; for model simulations it
is adopted $e$ = 0.5; \item the radial $V_w$ and azimuthal $U_w$
velocity components of the disk wind of the secondary (the
Keplerian velocity of the secondary in the periastron is assumed
to be equal to the unity: $V_k = 1$); \item the mass loss rate
from the accretion disk of the secondary $\dot M_w$.
\end{itemize}

In calculations of the optical characteristics of the dust
component in the disk wind we adopted the same dust to gas ratio
as in the interstellar medium 1:100. For simplicity we consider a
mono-dispersed graphite and silicate mixture of the dust particles
whose chemical composition is analogous to that in the
interstellar medium (Mathis et al. 1977). We used so-called
astrosilicate in our calculations. The radius of the particles $s=
0.1\mu$m, the average density is equal to 3 g cm$^{-3}$. The
optical characteristics of such particles were calculated with the
Mie theory in Paper I.

\section{Results}
We calculated a series of the common envelope models for the
different phases of the orbital period with the help of the method
described above. As an example, one of them is presented in Fig.
2. In the binaries with the elliptic orbits a concentration of the
particles on the line-of-sight in the direction towards the
primary (hereafter we shall call this parameter as a column
density of the probe particles $N$) depends not only on the phase
of the orbital period $\phi$ and an inclination of the orbit plane
to the line-of-sight $\theta$ but also on an orientation of the
orbit relatively to an observer. Fig 3. shows four variants of the
orientation of the secondary's orbit relatively to the observer
for which the column densities $N$ were calculated as a function
of $\phi$ as well as corresponded optical depths $\tau$.
\begin{figure*}
\hspace{3cm}
\includegraphics[width=10cm]{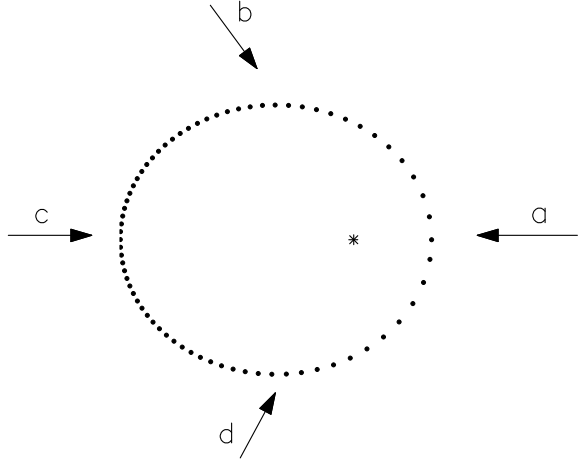}
\caption{The orbit of the
low mass secondary companion of the binary with an eccentricity $e
= 0.5$. Arrows point out the four different directions to the
observer for which the phase dependence of the column density in
Fig. 4 and the light curves in Figs. 5-7 have been calculated.}
\end{figure*}
A transition from $N$ to the column density of the real particles
$N_d$ was fulfilled using re-scaling when the ratio of the total
number of the dust particles ejected by the wind over one
revolution to the corresponded number of the probe particles was
taking into consideration (see Paper I for details). Also we took
into account a difference in the cross-sections of the columns:
the column density of the probe particles was calculated per the
section $\sigma$ (see below), while that of the dust grains $N_d$
was normalized to the cross-section of 1 cm$^2$.

As an example, Fig. 4 demonstrates a dependence of $N$ on $\phi$
for one of the model considered. The integer values of the phase
correspond to the moments when the secondary passes the
periastron. Calculations were made for the different angles of the
inclination of the orbit plane $\theta$ and the different
orientations of the orbit relatively to the observer. The
cross-section of the column $\sigma$ was adopted to be equal to
0.1\,$a$\, x\, 0.2$\,a$, where $a$ is the large semi-axis of the
secondary's orbit. The test calculations showed that for less
values of $\sigma$ the statistic fluctuations increased because of
the low number of the probe particles in the column while for
larger values of $\sigma$ smoothing of the details on the light
curves took place.
\begin{figure*}
\vspace{-1.1cm}
\includegraphics[angle=-90,width=14cm]{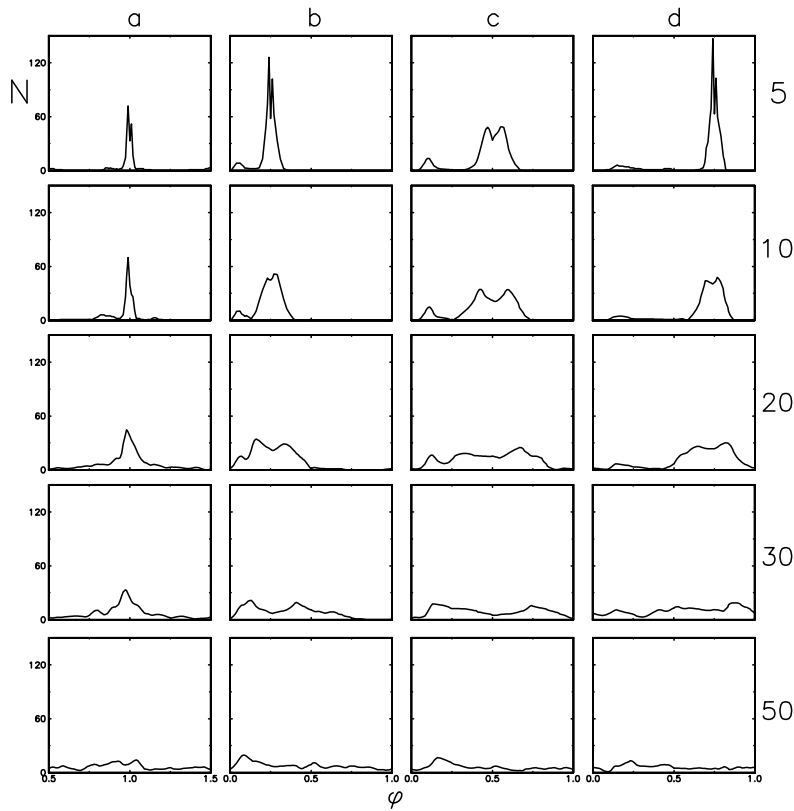}
\caption{A dependence of the column density of the probe particles
$N$ on the phase of the orbital period $\phi$ in the model with
$V_w = 1, U_w = 0, e = 0.5$ for four orientations of the binary
system relatively to the observer. The angle $\theta$ between the
line-of-sight and the equatorial plane of the system is given on
the right side.}
\end{figure*}

\subsection{Amplitudes and shapes of the light curves}
As in Paper I, when calculating a dilution of the light of the
binary due to the variations of the extinction on the
line-of-sight, we adopted that the main source of the radiation
was the main component which was treated as a point source. The
radiation flux from it decreases when going through the dust
component of the disk wind by $e^{-\tau}$ times. When $\tau \gg
1$, that means a strong dilution of the direct radiation of the
primary, the radiation flux is determined by the scattered
radiation of the CS dust included that of the CS and CB disks as
well as the common envelope. Taking this into account we can write

\begin{equation}
F_{obs} = \frac{L_*}{4 \pi D^2}\, e^{-\tau} + F_{sc}\,,
\end{equation}
where $D$ is a distance from the observer.

As a rule, a scattered light gives only a weak deposit to the
direct radiation of the young star and its main function is to
restrict the amplitude of the minima in those cases when the
direct radiation of the star is strongly diluted due to the
absorbtion in the CS medium. Such limiting functions of the
scattered radiation of the CS disks of the young stars are the
well known phenomenon in the case of UX Ori type stars whose
brightness undergoes strong decreases due to the variable CS
extinction (Grinin 1988).

In Figs. 5-7 a series of the light curves calculated for the
different models is presented. For simplicity it is adopted that
the flux of the scattered radiation does not depend on the phase
of the orbital period and is equal to 0.1\,$F_*$. Calculations are
made for the wavelength 5500\AA \hspace{0.3mm}  centered to the
V-band.

Like in the models with the circular orbits (Paper I), the light
curves in Figs. 5-7 have an asymmetric two-component shape caused
by a conic structure of the disk wind. At some system orientations
the two-component shape of the minima is not revealed due to the
low resolution because of the column cross-section $\sigma$
adopted in calculation of $N(\phi)$. Since the stellar disk of the
main companion has finite sizes, a similar smoothing of the light
curves can also occur in the real situations.
\begin{figure*}
\vspace{-1.5cm} \hspace{1cm}
\includegraphics[angle=-90,width=13cm]{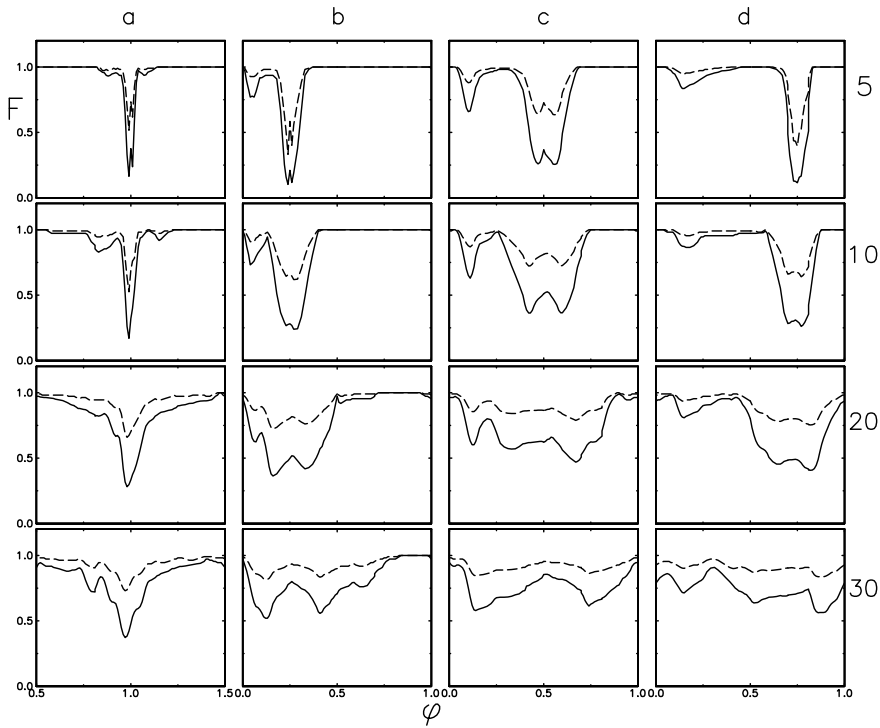}
\caption{Theoretical light curves of the main system component for
the orbit orientations shown in Fig. 3. Model parameters are $e =
0.5, V_w = 1, U_w = 0$, the mass loss rate of the disk wind of the
secondary $\dot{M_w}=10^{-8} M_\odot$ yr$^{-1}$ (solid line) and
$3 \cdot 10^{-9} M_\odot$ yr$^{-1}$ (dashed line). The angle
$\theta$ between the line-of-sight and the plane of the double
system is given on the right side.}
\end{figure*}

One sees from Figs. 5-7 that at the same other conditions, the
longest minima take place in the systems whose eclipses are
happened at the moment of passing the apastron by the secondary,
and to the contrary, the shortest eclipses has to be observed
from the side of the periastron. It is easy to show that a
ratio of the eclipse durations at these opposite orientations of
the binary is

\begin{equation} \Delta t_a/\Delta t_p = (1+e)/(1-e)\,.
\end{equation}
Hence it follows, that when $e$ = 0.5 durations of eclipses differ
by 3 times.

In the most models considered at the orbit orientation $b$ (Fig.
3) (when an eclipse occurs after a passage of the periastron), the
light curve is characterized by the steep descent and a slow
ascent while in the same models but in the position $d$ (when an
eclipse occurs before a passage of the periastron) the result is
the opposite: a decrease of the brightness is slower compared to
its increase.

At the small inclination angles of the orbit a duration of the
eclipses is less compared to the period of the orbital motion. An
exception are the systems oriented with the apastron to the
observer (the variant $c$ in Figs. 5-7): in such cases, even under
a small $\theta$ an eclipse can last for a rather long time. With
increase of the orbit inclination a duration of the eclipses
increases and for $\theta \ge 30^\circ$ can be compared with the
system period.

As it is seen from Fig. 5, in the models with the low velocities
of the disk wind one can see a weaker minimum preceding the main
one. A similar feature is also present in the models with circular
orbits. As shown in Paper I, its origin is caused by the wind
particles ejected during the previous passage of the secondary
component on the orbit and captured by the main component. In the
models with the high mass loss rate ($\dot M_w \ge 10^{-7}
M_{\odot}$ yr$^{-1}$) and the large inclination angles of the
orbit a boundary between the main minimum and its predecessor
vanishes resulting the more extended two-component structure of the minima.
In the models with larger wind velocity (Fig. 6) a
portion of the matter captured by the primary decreases resulting
only one main minimum on the light curves.
\begin{figure*}
\vspace{-1.5cm}
  \includegraphics[angle=-90,width=15cm]{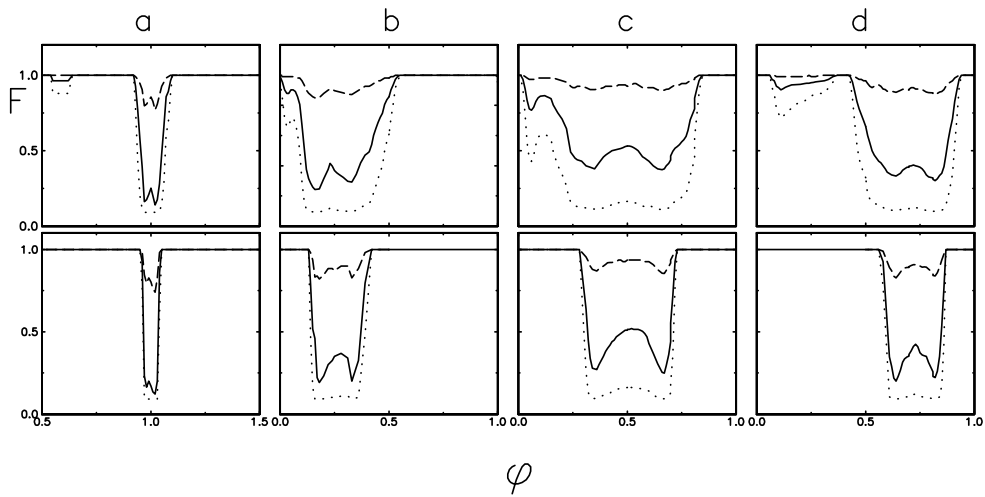}
  \vspace{-2cm}
  \caption{The light curves for the model with parameters: $V_w = 2,
  U_w = 0, e=0.5$ with the mass loss rate $\dot M_w = 3 \cdot 10^{-9}
  M_\odot$ yr$^{-1}$ (dashed line), $\dot M_w = 3 \cdot 10^{-8} M_\odot$
  yr$^{-1}$ (solid line) and $\dot M_w = 10^{-7} M_\odot$ yr$^{-1}$
  (dots). The angle $\theta = 20^\circ$. Top: $40 \le \alpha \le 60^\circ$,
  bottom: $\alpha = 45^\circ$}
  \end{figure*}

Calculations show that taking into account a rotation of the disk
wind yields a more rapid expansion of the wind in the horizontal
directions in comparison with the models where $U_w = 0$. As a
result, the depth of the minima decreases but their duration
increases (Grinin and Tambovtseva 2003). The minima parameters and
their shape depend also on the geometry of the disk wind. Fig. 6
demonstrates results of calculations for two models differing only
with open angle of the wind; in one case it is the same as in Fig.
5 ($40 \le \alpha \le 60^\circ$), in another case it is
45$^\circ$. It is seen that in the last case the minima are
shorter and have a more symmetric shape.

We remind that all model presented above are calculated in
suggestion of a constant mass loss rate. For comparison we also
considered the models with a variable mass loss rate: $\dot M_w
\propto \dot M_a$ in which a dependence of $\dot M_a$ on $\phi$
has been taken from AL96. Calculations showed (Fig. 7) that in the
models with the variable $\dot M_w$ an amplitude of the minima
depended on the orientation of the orbit of the secondary
relatively to the observer that is quite natural: it was maximal
in the systems whose periastron was between the observer and the
primary (the variant $a$ in Fig. 3) and minimal in the case of the
opposite orbit orientation (the variant $c$ in Fig. 3). As for a
shape of the light curves, the difference between these two models
is not so essential.
\begin{figure*}
\vspace{-2cm}
  \includegraphics[angle=-90,width=15cm]{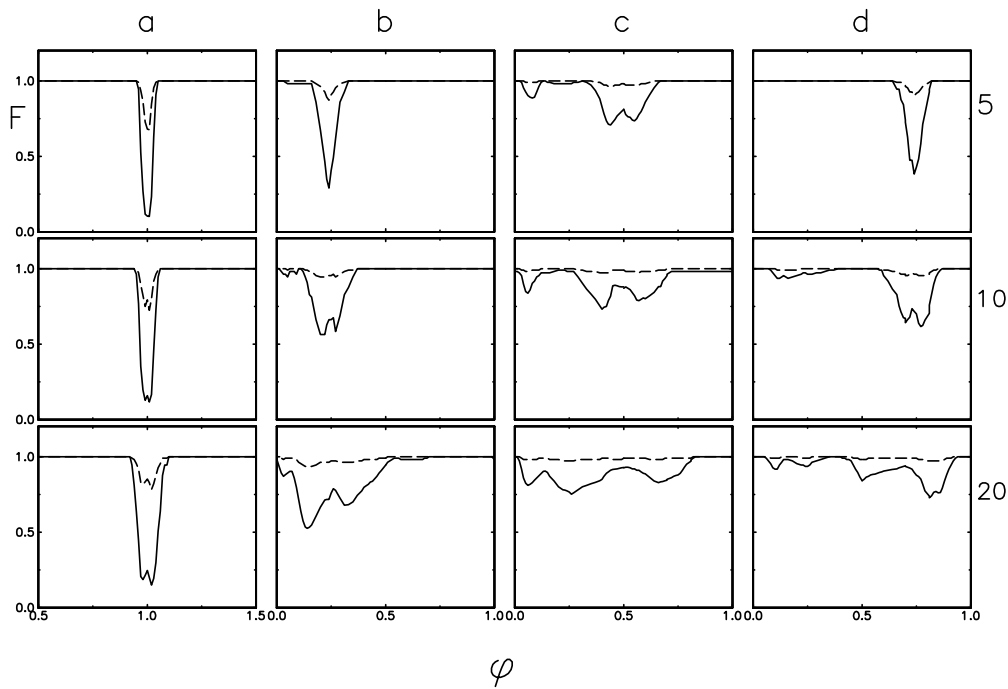}
  \vspace{-2cm}
  \caption{Light curves for the model with a variable mass loss
  rate ($\dot M_w \propto \dot M_a$). The solid line corresponds
  to $\dot M_w = 3\cdot 10^{-8} M_\odot$ yr$^{-1}$ in the
  periastron and the dashed line to $\dot M_w = 3\cdot 10^{-9}
  M_\odot$ yr$^{-1}$ in the periastron. The model parameters are
  $V_w = 2, U_w = 0, e = 0.5$. The angle $\theta$ between the
  line-of-site and the double system plane is given on the right
  side.}
  \end{figure*}

\subsection{Modulation of the scattered radiation with the phase of the
orbital period} When calculating light curves, we adopted for
simplicity that an intensity of the scattered radiation did not
depend on the phase of the orbital period. Such a situation is
possible if the CB disk is a main source of the scattered
radiation. In our case the part of the scattered radiation
originates from the disk wind of the secondary and the common
envelope created by it, and this part undergoes a periodical
modulation. Let us consider a simple example that demonstrates how
the phase dependence of the scattered radiation can influence the
light curves of the binary.

For this purpose, the intensity of the scattered radiation for two
of models considered has been calculated in a single scattering
approach. (An analysis showed that this approach was valid for the
mayor part of the common envelope at $\dot M_w \le 10^{-8}
M_{\odot}$\, per year). The flux of the scattered radiation in
this approach is
\begin{equation} F_{sc} = \frac{L_*}{4\pi D^2}\, \pi
s^2\,Q_{sc}\, \int_{V}{n({\bf r})} \,r^{-2}\, f(\gamma)\,e^{-{\bf \tau}_1-{\bf
\tau}_2}\,d {\bf r}
\end{equation}
Here ${n(\bf r)}$ is a concentration of the particles at the point
${\bf r}$, $Q_{sc}$ is a scattering efficiency factor of the dust
grain, $s$ is its radius, $\tau_{1}$ is an optical depth between
the point ${\bf r}$ and the primary, ${\tau_{2}}$ is an optical
depth between the point ${\bf r}$ and the observer. The
integration in Eq. (7) is carried out over all volume $V$ of the
common envelope.

As in the previous section, we assume that the dust consists of
the mixture with equal amounts of the graphite and astrosilicate
particles with the radius $s = 0.1 \mu$m. A scattering process is
described with the Henyey-Greenstein phase function
\begin{equation}
f(\gamma) = \frac{1}{4\,\pi}\frac{1-g^2}{(1+g^2-2g\,\cos{\gamma})^{3/2}}
\end{equation}
where $\gamma$ is a scattering angle, the asymmetry factor $g$ is
assumed to be equal to 0.5.

Fig. 8a demonstrates the results of calculations for the two
system orientations: almost edge-on ($\theta = 5^\circ$) and
pole-on. In both cases an integer value of the phase $\phi$
corresponds to the moment of the periastron passage by the
low-mass companion. In the first case the periastron is between
the primary and the observer. It is seen that the flux of the
scattered radiation $F_{sc}$ reaches its maximum namely in this
phase. When observing the system edge-on, this takes place due to
the forward elongated scattering phase function. When observing
the system pole-on, a maximum of $F_{sc}$ is reached due to that a
dilution coefficient weighted over all envelope is maximal at the
moment of the periastron passage by the low-mass companion.

One can see from Fig. 8a that a dependence of $F_{sc}$ on the
orbital period phase is asymmetric and characterized with a steep
growth when the secondary approaches the periastron and a slower
decrease when the secondary moves away from it. Calculations show
that a degree of asymmetry depends both on the value of the
dimensionless velocity of the disk wind $V_w$ and on the
eccentricity of the orbit $e$ and its inclination to the
line-of-sight. In the model considered above (Fig. 8a) a maximal
asymmetry of $F_{sc}$ is reached if the system is observed
pole-on.

The functions $F_{sc}$ of $\phi$ given in Fig. 8a were calculated
for the model with $\dot M_w = const $. In the models with $\dot
M_w \propto \dot M_a$ a total picture is qualitatively the same.
As earlier, a maximum of the scattered radiation is reached in the
periastron of the orbit and has even larger amplitude.
\clearpage
\begin{figure*}
\hspace{-2cm}\includegraphics[width=18cm]{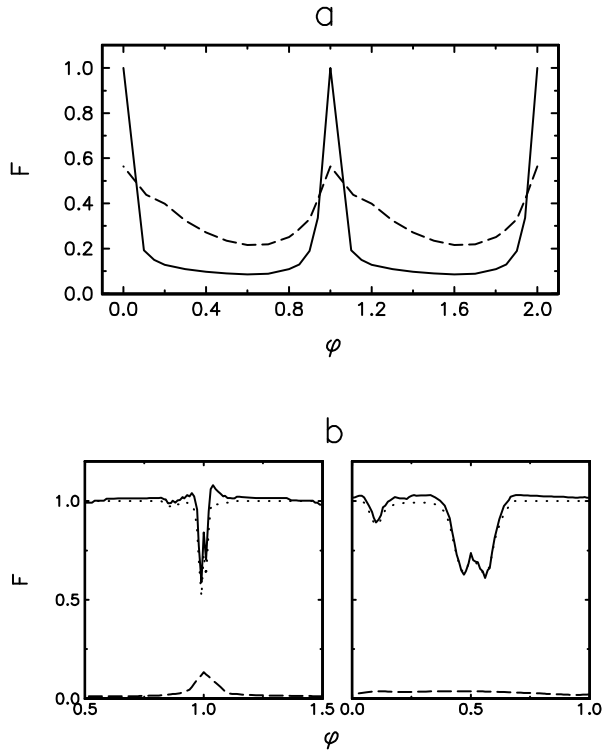}\\
\vspace{-2cm}
  \caption{a: An example of the phase dependence of the scattered
  radiation flux for two system orientations $\theta = 5^\circ$
  (solid line) and $\theta = 90^\circ$ (dashed line). The model parameters
  are $V_w = 1, U_w = 0, e = 0.5$;
  b: The model of the eclipse of the
  main companion in the binary system with the account of the phase
  dependence of the scattered radiation for the same
  model: in the periastron (left) and apastron (right). The dots
  indicate to the light curves without the scattered
  radiation. The dashed line shows a behavior of the scattered light
  flux with phase.}
  \end{figure*}

If to use the values of $F_{sc}$ obtained above in Eq. (5), then
asymmetric low-amplitude brightening appears on the light curve on
going and/or outgoing the star from the minimum (Fig. 8b). A
similar details are observed from time to time on the light curves
of some eclipsing young objects (see, e.g. G\"urtler et al. 1999;
Herbst et al. 2002) and can occur due to scattering the radiation
of the primary by those dust particles who have a forward
scattering phase function. It is also obviously that taking into
account a phase dependence of the scattered light one can expect a
brightness increase in the central part of the deep minima whose
amplitude is restricted from below to a scattering radiation.

\subsection{FUOR-like light curves}
As calculations showed, in the models with the large mass loss
rate a common envelope formed by the disk wind of the secondary
can be so powerful that during a substantial part of the orbital
cycle the main component can be obscured from an observer. In such
cases the light curve looks like a series of the subsequent
flares; the latter, in fact, represents a short time intervals
corresponding to the minimal values of the extinction on the
line-of-sight (Fig. 9).

This result can be interesting in connection with the debates on
the nature of FUORs (see Herbig et al. 2003 and
the literature cited there). According to Hartmann and Kenyon
(1985) the outbursts of these objects are caused by the repeated
"outbursts" of the accretion rate onto the young stars resulting
an increase of the luminosity of the accretion disk that is
associated with the FUOR flare.
\begin{figure*}
  \includegraphics[angle=-90,width=14cm]{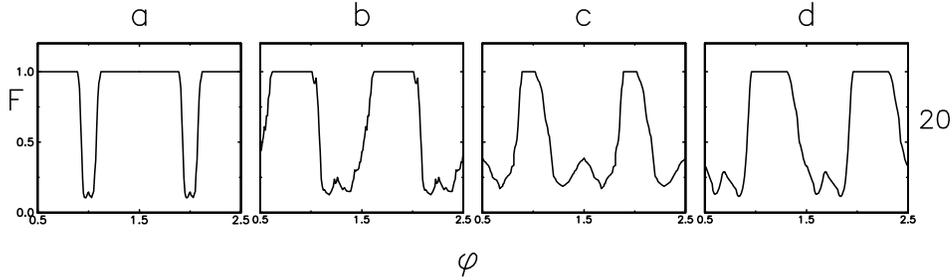}
  \vspace{-6cm}
  \caption{The light curves for the mass loss rate $\dot M_w=10^{-7}
  M_\odot$ yr$^{-1}$ in the model with parameters $V_w = 3, U_w = 0.5,
e = 0.5$ and $\theta = 20^\circ$}
\end{figure*}
According to another hypothesis (Herbig et al. 2003), the FUORs
outbursts are caused by the changes in the star luminosity. Both
hypotheses are not free of difficulties. Therefore, as an
alternation, it is interesting to consider a possibility to
interpret the light curves of some FUOR-like objects on the base
of the young binary model, in which the primary (a star of a high
luminosity) is periodically screened by the gas and dust common
envelope formed by the disk wind of the secondary.

\section{Discussion}
Thus, the theoretical light curves of the young binaries in which
the eclipses of the main component are caused by the disk wind of
the secondary companion are characterized with a large variety of
the shapes, and in many cases absolutely do not resemble the
classical light curves observed in the usual eclipsing binaries.
If the mass loss rate by the accretion disk of the secondary $\dot
M_w \ge 3\,10^{-8}$, the eclipses can be observed even if its
orbital plane is strongly inclined to the line-of-sight. The
duration of such eclipses can be compared with a period of the
orbital motion. Taking into consideration this fact as well as a
large number of the double and multiple systems among the young
stars (see the review by Mathieu et al. 2000), one can expect an
existence of the large-scaled cyclic brightness variability in
many young stars. Therefore, it is reasonable to suppose, that
several young eclipsing systems with the long lasting eclipses
discovered not long ago is, in fact, a rather wide-spread
population of the young stars whose number will be rapidly
increase with an accumulation of the photometric data on the young
clusters. We suppose that the UX Ori type stars also belong to
such eclipsing systems; the cyclic activity of these stars have
been interpreted in assumption of their binarity by Grinin et al.
(1998), Rostopchina et al. (1999) and Bertout (2000).

There is one more feature in the model of eclipse considered here
which is necessary to take into account for an analysis of the
continuous photometric observations of the young stars. As
mentioned in Section 2, in the binaries with the eccentric orbits
the CB disk is characterized with the global asymmetry and slowly
precesses. Observations of such systems under a small inclination
to their equatorial plane have to reveal slow (secular) variations
of the extinction and, hence, the brightness of the primary.
Theoretically a situation is possible, when a direct radiation of
the companions will be completely blocked by the CB disk during
some time. In addition, under the effects of the tidal
interactions with the CB disk, the orbits of the companions also
have to precess slowly. This means, that in the models of eclipses
considered a slow transition from the variant of the eclipse $a$
to $d$ (Figs. 5 - 7) has to take place implying slow changes in
the parameters of minima: their depth, duration and shape.

Possibly, the exotic eclipsing system KH 15D mentioned in
Introduction belongs to such objects; its binarity has been
revealed recently by Johnson et al. (2004) on the base of the
variations of the radial velocity of the main component and
estimated eccentricity (0.68 $\le e \le$ 0.80). An analysis of the
archive photographic observational data (Johnson and Winn 2004)
showed that the brightness of KH 15D was slowly decreasing during
the last 40 years (see also Barsunova et al. 2004 on this topic).
The duration of eclipses (Herbst et al. 2002) and their depth
(Johnson and Winn 2004) were also slowly changing. According to
the data of the Harvard collection of the photographic
observations the duration of the KH 15D eclipses in the beginning
of the last century was essentially less than at the present time
(Winn et al. 2003). All these features can be explained with the
disk wind eclipse model at the condition that the orbit of the
secondary is highly elongated and precesses.

Another possible application of the model considered can be
connected with an interpretation of the large scale photomertic
activity of some candidates to FUORs. An analysis of the published
photometry of such objects as V1515 Cyg (Hartmann et al. 1993) and
Z CMa (van den Ancker et al. 2004) shows that their light curves
look like the theoretical ones presented in Figs. 5-7. Keeping
this in mind and that some FUORs are members of the double
systems, one can consider a possibility of the substantial (in
amplitudes) variations of their brightness due to the cyclic
variations of the extinction on the line-of-sight as reliable and
worth of further investigations\footnote{Note, that Bonnel and
Bastien (1992) discussed the role of the FUORs binarity in
connection with a possibility of the periodic increase of the
accretion rate onto the main companion initiated the flares to
their opinion.}

As we saw above, a dependence of the scattered radiation on the
phase of the orbital cycle can lead to appearance of the fine
details on the light curves: a small brightness increase when
ongoing and/or outgoing the minimum. More essential, however, the
circumstance that the motion of the densest part of the common
envelope on the orbit has to evoke a periodical modulation both of
the intrinsic polarization of the young binary and its infrared
excess. And this modulation can be observed at any orientation of
the binary relatively to the observer. This gives a possibility to
reveal close binaries whose orbit planes are close to the sky
plane and which are hardly discovered for this reason.

At last, it should be noted that one of the assumption made for
calculation simplifying was about a negligible influence of the
primary disk wind on the common envelope structure. Such an
approach was valid in the model of binary considered above since
the secondary was the principle accretor and, therefore, the main
source of the matter forming the common envelope. However, even at
this condition, the role of the primary disk wind can be essential
when investigating the cycle activity of the binary at the other
wavelengths, and first of all, in the X-ray. Really, as estimates
show, a collision of the two disk winds even with rather modest
mass loss rates of about of 10$^{-9}M_\odot$ and the velocity of
100 km/s is equivalent to the energy release of order of 3$\cdot
10^{30}$ erg/s. It is comparable with the X-ray luminosity of many
young stars (see the review by Glassgold et al. 2000) and has to
be taken into account in interpretation of the X-ray radiation of
the young objects.

\section{Conclusion}
The results presented above show that the behavior of the optical
characteristics of the young binaries with the eccentric orbits
strongly depend on that how two main processes are connected: an
accretion of the matter from the CB disk onto the system
components and a matter outflow from the accretion disks
surrounding them. Therefore, one of the clue task in the physics
of the young binaries is a self-consistent solution of the problem
when both of these processes are tightly connected. In particular,
it would be important to investigate the role of the tidal
interactions in the formation of the disk winds of the system
companions and determine a phase dependence of the mass loss rate
from the accretion disk of the secondary component. \\
\\
The authors are thankful to P. Artymowizc for the useful
discussion of the problems touched in the paper as well as to S.A.
Lamzin who read the manuscript and made valuable comments. The
work has been supported by grant INTAS 03-51-6311, grant of the
Presidium of the Russian Academy of Sciences "Non-stationary
phenomena in Astronomy" and grant of Scientific School
1088.2003.2.

\Large
\begin{center}
{\bf References}
\end{center}
\normalsize
Artymowicz P., Lubow S.H., 1994, ApJ {\bf 421}, 651\\
Artymowicz P., Lubow S.H., 1996, ApJ {\bf 467}, L77\\
Barsunova O. Yu., Grinin V. P. and Sergeev S., 2004, in preparation\\
Bate M.R., Bonnell I.A., 1997, MNRAS, {\bf 285}, 33\\
Bertout C., 2000, A\&A , {\bf 363}, 984\\
Bisikalo D.B., Boyarchuk A.A, Kuznetsov O.A. et al. 1995,
Astronomy Reports {\bf 72}, 190\\
Cohen R.E., Herbst W., Willams E.C., 2003, astro-ph/0308484\\
Dequennoy A., Mayor M., 1991, A\&A, {\bf 248}, 485\\
Glassgold A.E., Feigelson E.D., Montmerle T., 2000, in "Protostars
and Planets IV", Eds. V. Mannings, A.P. Boss and S.S. Russel,
(Univ. Arizona Press, Tucson), p.429\\
Goodson A.P., B\"ohm K.-H., Winglee R., 1999, ApJ {\bf 524}, 142\\
Grinin V.P., 1988, Astronomy Let. {\bf 14}, 65\\
Grinin V.P. 2002, Astronomy Rep. {\bf 46}, 380\\
Grinin V.P., Tambovtseva L.V. 2002, Astronomy Let. {\bf 28}, 601\\
Grinin V.P., Tambovtseva L.V., Proc. "Towards Other Earths:
DARWIN/TFP and Search for Extrasolar Terrestrial Planets", 2003,
Heidelberg, Germany, Eds. M. Fridlund, T. Henning, ESA SP-539, p.
429\\
Grinin V.P., Rostopchina A.N., Shakhovskoi D.N., 1998, Astron.
Reports, {\bf 24}, 802 \\
G\"urtler J., Friedemann C., Reimann H.-G., et al. ), 1999,
Astron. Astrophys. Suppl. {\bf 140}, 293\\
Hamilton, C.M., Herbst W., Shih C., Ferro A.J., 2001, ApJ {\bf 554}, L201\\
Hartigan P., Edwards S.E., Ghandour L., 1995, ApJ {\bf 436}, 125\\
Hartmann L., Kenyon S.J., 1985, ApJ {\bf 299}, 462\\
Hartmann L., Kenyon S., Hartigan P., 1993, in {\it Protostars and
Planets} III, Eds. E.H.Levy and J.I.Lunine, (Univ. Arizona Press,
Tucson) p. 497\\
Herbig G.H., Petrov P.P. Duemmler R., 2003, ApJ, {\bf 595}, 348\\
Herbst W., Hamilton C.M., Vrba F., et al., 2002, PASP, {\bf 114},
1167\\
Hernquist L., Kats H., 1989, ApJSS, {\bf 70}, 419,\\
Hirth G.A., Mundt R., Solf J., 1997, Astron. Astrophys. Suppl.
{\bf 126}, 437\\
Ismailov N.Z., 2003, Astron. Rep. {\bf 47}, 206\\
Johnson J.A., Marcy G.W., Hamilton C.M. et al., 2004, Astro-ph 0403099\\
Johnson J.A., Winn J.N., 2004, Astro-ph 0312428\\
Kearns K.M., Herbst W., 1998, AJ {\bf 116}, 261\\
Lubow S.H., Artymowicz P., 2000, in {\it Protostars and Planets
IV}, Eds. by V. Mannings, A.P. Boss and S.S. Russel, (Univ.
Arizona Press,
Tucson) p. 731\\
Makita M., Miyawaki K., \& Matsuda T., 2000, MNRAS, {\bf 316} 906\\
Mathieu R.D., Adams F.C., Latham D.W., 1991, AJ, {\bf 101}, 2184\\
Mathieu R.D., Ghez A.M., Jensen E.K.N., Simon M., 2000, {\it
Protostars and Planets} IV, Ed. by V.Mannings, A.Boss,
S.S.Russell, (Univ. Arizona Press, Tucson),
p. 559\\
Mathis J.S., Rumpl W., Nordsieck K.H., 1977, ApJ {\bf 217}, 425\\
Mayor M., Urdy S., 2000, in {\it Disk, Planetezimals, and
Planets}, Eds. F.Garzon et al. ASP Conf. {\bf 219}, 441\\
Mazeh T., Goldberg D., Dequennoy A., Mayor M., 2000, ApJ, {\bf 401}, 265\\
Natta A., Grinin V.P., Mannings V. 2000, {\it Protostars and
Planets} IV. Ed. by V.Mannings,
A.Boss, S.S.Russell, (Univ. Arizona Press, Tucson), p. 559\\
Pogodin M.A., Miroshnichenko A.S., Tarasov A.E. et al., A\&A, in press \\
Rozyczka M, Laughlin G., 1997, in ASP Conf. Ser. 121,
{\it Accretion phenomena and related outflows}, 792\\
Rostopchina A.N., Grinin V.P., Shakhovskoi D.N., 1999, Astron.
Let. {\bf 25}, 243\\
Safier P.N., 1993, ApJ {\bf 408}, 115\\
Sawada K., Matsuda T., \& Hachisu I., 1986a, MNRAS, {\bf 219}, 75\\
Sawada K., Matsuda T., \& Hachisu I., 1986b, MNRAS, {\bf 221}, 679\\
Stone J.M., Gammie C.F., Balbus S.A., Hawley J.F., 2000, in {\it
Protostars and
Planets}. IV. Eds. V.Mannings, A.P.Boss, S.S.Russel, p. 589\\
Shakura N.I., Sunyaev R.A., 1973, A\&A, {\bf 24}, 337\\
Shevchenko V.S., Grankin K.N., Melnikov S. Yu., Lamzin S.A., 1998,
Astronomy Let. {\bf 24}, 528\\
van den Ancker M., Blondel P.F.C., Tjin A Djie H.R.E. et al. 2004,
astro-ph 0401338\\
Winn J.N., Garnavich P.M., Stanek K.Z., Sasselov D.D., 2003, ApJ,
{\bf 593}, L121\\
\end{document}